\documentstyle[bbm,amsfonts,epsfig,12pt]{article}

\newcommand {\eqref} [1] {(\ref {#1})}
\newcommand {\slsh} [1] {\not{\hbox{\kern-2pt${#1}$}}}
\newcommand{\drawsquare}[2]{\hbox{%
\rule{#2pt}{#1pt}\hskip-#2pt
\rule{#1pt}{#2pt}\hskip-#1pt
\rule[#1pt]{#1pt}{#2pt}}\rule[#1pt]{#2pt}{#2pt}\hskip-#2pt
\rule{#2pt}{#1pt}}
\newcommand{\Yfund}{\raisebox{-.5pt}{\drawsquare{6.5}{0.4}}}
\newcommand {\beq} {\begin{equation}} 
\newcommand {\eeq} {\end{equation}}
 \newcommand {\ber}{\begin{eqnarray*}}
 \newcommand {\eer} {\end{eqnarray*}}
\newcommand {\bea}{\begin{eqnarray}}
 \newcommand {\eea} {\end{eqnarray}} 
\def\Acknowledgements{\bigskip  \bigskip {\begin{center} \begin{large}
             \bf ACKNOWLEDGEMENTS \end{large}\end{center}}}

\begin{document}
\begin{titlepage}
\rightline{CERN-TH/2002-058}
\rightline{AEI-2002-023}
\vskip 1cm
\centerline{{\Large \bf Nonplanar Anomalies in Noncommutative Theories}}
\centerline{{\Large \bf and the Green-Schwarz Mechanism}}
\vskip 1cm
\centerline{Adi Armoni\ ${}^\dagger$, Esperanza Lopez\ ${}^\ddagger$
  and Stefan Theisen\ ${}^\ddagger$}
\vskip 0.5cm
\centerline{${}^\dagger$ Theory Division, CERN}
\centerline{CH-1211 Geneva 23, Switzerland}
\vskip 0.3cm
\centerline{adi.armoni@cern.ch}
\vskip 0.5cm
\centerline{${}^\ddagger$ Max-Planck-Institut f{\"u}r
  Gravitationsphysik, Albert-Einstein-Institut,}
\centerline{Am M{\"u}hlenberg 1,D-14476 Golm, Germany}
\vskip 0.3cm
\centerline{lopez,theisen@aei-potsdam.mpg.de}
\vskip 1cm

\begin{abstract}
We discuss nonplanar anomalies in noncommutative gauge
theories. In particular we show that a nonplanar anomaly exists when
the external noncommutative momentum is zero and that it leads to a
non-conservation of the associated axial charge. In the case of
nonplanar local anomalies, a cancellation of the anomaly can be
achieved by a Green-Schwarz mechanism. In an example of D3
branes placed on an orbifold singularity that leads to a chiral
theory, the mechanism involves twisted 
RR fields which propagate with zero noncommutative momentum.
Global anomalies are not cancelled and, in
particular, the decay $\pi^0 \rightarrow 2\gamma$ 
is allowed.
\end{abstract}
\end{titlepage}

\section{Introduction}

Anomalies play a fundamental role in particle physics
 \cite{Treiman:ep}. In field
theory, local anomalies signal an inconsistency of the theory and are
therefore considered as `bad'. Global anomalies
are `good' since they allow decay processes, such as $\pi ^0
\rightarrow 2\gamma$, which are classically forbidden but observed 
in nature.   
Through anomaly matching conditions, global
anomalies restrict microscopic and effective descriptions \cite{tHooft}
which was important in establishing Seiberg duality \cite{Seiberg:1994pq}.
In a consistent string theory, all (local) anomalies must cancel.
The manifestation of the consistency 
in the low energy effective theory is, however, subtle, as it might involve
a cancellation between a tree level diagram and a one loop diagram.
This is the Green-Schwarz mechanism \cite{Green:sg}.

Recently, there has been much interest in noncommutative field theories,
see \cite{Douglas:2001ba,Szabo:2001kg} for reviews.
The issue of anomalies in the framework of noncommutative chiral
gauge theories was already discussed by several authors
\cite{Ardalan:2000cy}--\cite{Banerjee:2001un}.
There are two types of possible anomalies: planar and nonplanar. Planar
anomalies are well understood. They result from graphs that `behave'
exactly as in the commutative theory apart from an overall phase
factor which depends only on external momenta
\cite{Gonzalez-Arroyo:1982ub,Filk:1996dm}, and
therefore automatically obey the usual anomaly 
equations, such as {\it e.g.} eq.(\ref{pl}) below. 

The issue of nonplanar anomalies is subtle. Nonplanar graphs
are UV finite at one loop and therefore it was suggested
that these anomalies automatically vanish 
\cite{Intriligator:2001yu,Bonora:2001fa,Martin:2001ye}. However,
Ardalan and Sadooghi
pointed out \cite{Ardalan:2000qk} that due to UV/IR mixing
\cite{Minwalla:1999px}, the nonplanar
anomaly does not vanish (see also \cite{Banerjee:2001un}).
\footnote{It is interesting to note that a
similar phenomenon of vanishing anomaly exists for the lattice
version of the noncommutative theory \cite{Nishimura:2001dq}. It
is, however, not clear whether it survives the continuum limit.}

The goal of this work is to clarify the issue of nonplanar anomalies. 
On one hand, if one views noncommutativity as a 
(gauge invariant) regulator, it is hard
to believe that one can introduce a regulator that leads to a
conservation of the axial charge (noncommutativity is, however, 
not a good regulator in the sense that the limit $\theta \rightarrow 0$ is not
smooth). Another argument against the vanishing of the anomaly arises
from ${\cal N}=2$ SYM \cite{Armoni:2001br}. In this theory the
$U(1)_{\cal R}$ anomaly belongs to the same multiplet of the conformal
anomaly (the $\beta$-function). Since the $\beta$ function,
generically, does not vanish, it would be surprising if the
$U(1)_{\cal R}$ anomaly vanishes.
On the other hand the paradox raised in 
\cite{Intriligator:2001yu} seems to suggest the vanishing of
the nonplanar anomaly as a unique resolution.

The paradox is the following: consider a $U(N)$ chiral theory which can be
realized by a brane configuration in string theory. Suppose that
the $U(1)$ is potentially anomalous, due to an excess of left handed
fermions over right handed fermions. In the ordinary theory, that
would lead to an anomaly, which can be resolved in string theory by giving 
an infinite mass to the $U(1)$, leaving at low energies an $SU(N)$
theory and a global $U(1)$. 
In those non-commutative gauge theories which have been obtained as 
effective low-energy field theories of string theory one always 
obtains $U(N)$ gauge groups, rather than $SU(N)$. In these theories
the noncommutativity mixes the $U(1)$ and the $SU(N)$ parts and  
the above scenario cannot occur
\cite{Matsubara:2000gr,Armoni:2000xr}. 
A way out, which was proposed in \cite{Intriligator:2001yu}
is that actually there is no anomaly, though the
matter content is left-right asymmetric, due to the finiteness of the
associated nonplanar graph.

We would like to present a different solution: 
the nonplanar anomaly graph is indeed finite and actually vanishes, for
any {\em non-zero noncommutative external momentum}. However, the
graph is not regulated for exactly zero external noncommutative
momentum - which leads to the integrated anomaly equation 
\beq
\int d^2x_{NC}\ \partial_\mu j^\mu_A=-{g^2\over 8 \pi^2}\int d^2x_{NC}\ 
F_{\mu\nu} \tilde F^{\mu\nu} \label{intanomaly}.
\eeq
In this expression, the integration is restricted to the noncommutative
plane\footnote{A similarly needed integration over the noncommutative
plane was pointed out \cite{Zamora:2000ce} in relation with
renormalization scale independence of composite operators.}.
In particular the anomaly does not vanish. As we shall see,
despite of the anomaly, string theory avoids the inconsistency via
a subtle Green-Schwarz mechanism, which involves closed strings with 
exactly zero noncommutative momentum. The resulting low-energy theory 
would be $U(N)$ (with $U(N)$ gauge invariance !), with the $U(1)$
being massive only if it carries zero noncommutative momentum.

We will also discuss global anomalies. Here we claim that due to
\eqref{intanomaly} the global anomaly does not vanish. In particular
the $\pi^0$ decays. In addition, if one considers anomaly matching
conditions, nonplanar anomalies should be taken into account as well.

The organization of the manuscript is as follows: in section 2 we
describe the ambiguity (or freedom) in the definition
of currents in the noncommutative theory and the relevance of non-planar
anomalies. In section 3 we review the arguments, using perturbation theory,
in favor of the vanishing of the anomaly for any $\theta q \neq 0$
and we show that the behavior at $\theta q=0$ leads to an integral 
anomaly equation. In section 4 we show that the point splitting definition 
of the non-planar current gives rise to the same integrated anomaly equation. 
We carry the analysis both for two and four dimensions. In section 5
we discuss, briefly, mixed anomalies. Section 6 is devoted 
to a discussion and a resolution of the above mentioned 
paradox in field theory and in string theory. Finally, in section 
7 we discuss the consequences of our findings on global anomalies.

We use the following conventions throughout the manuscript.
$[x^\mu,x^\nu]_\ast = i\theta ^{\mu \nu}$ and $\theta q$ stands for $\theta
^{\mu \nu} q _\nu$. In 4d we consider only theories with space-space
 noncommutativity and in 2d we assume an Euclidean signature. 
Frequently we will use the notation 
$F(q)=f(q)|_{\theta q=0}$. By this we mean that $F(q)=f(q)$ only if $q$ has a
zero component along the noncommutative directions, otherwise
$F(q)=0$.

\section{Anomalies in Noncommutative Gauge Theories}

We will consider a non-commutative gauge theory with group 
$U(1)$ and a massless Dirac fermion transforming in the 
fundamental representation. The fermionic action is
\beq
S= i \int d^d x \, {\bar \psi} \, \ast  \slsh{\! D} \psi \, ,
\eeq
where $D_\mu \psi\!=\! \partial_\mu\psi \!+\! ig A_\mu \! \ast\! \psi$. This
action is invariant under global axial transformations
\beq
\delta_\alpha \psi(x)=i \, \alpha \,\gamma^5 \psi(x) \, .
\label{axial}
\eeq
In trying to derive the axial associated current we are faced with a 
problem \cite{Gracia-Bondia:2000pz,Ardalan:2000qk}. If we define 
the axial current following the Noether procedure, we have to decide 
if the lagrangian is given by {\it i}) ${\cal L}=i{\bar \psi} \, \ast  
\slsh{\!\! D} \psi$ or instead {\it ii}) ${\cal L}'=-i(\slsh{\!\! D} \psi)^t 
\ast {\bar \psi}^t$
\footnote{Of course one can imagine an intermediate definition 
interpolating among the two previous possibilities.}. 
The two lagrangians lead to the same action since under integration
the $\ast$-product satisfies cyclic symmetry. Equivalently, in order
to derive the axial current we can formally promote 
$\alpha \rightarrow \alpha(x)$. Then we should choose 
$\alpha(x)$ to multiply $\psi(x)$ on {\it i}) the right or {\it ii})
the left. The axial currents associated to these two choices are 
\beq
i) \;\; j^\mu_A = {\bar \psi} \ast \gamma^\mu \gamma^5 \psi 
\;\;\; , \;\;\;\;\;\;
ii) \;\; j'^{\, \mu}_A =- \psi^t (\gamma^\mu \gamma^5)^t \ast {\bar \psi}^t 
\, . \label{currents}
\eeq

The conservation properties of these two currents have been extensively
studied, giving different results for $j_A$ and $j'_A$ 
\cite{Ardalan:2000cy}--\cite{Martin:2001ye}. This apparently
puzzling conclusion has the following origin. Although the lagrangians
${\cal L}$ and ${\cal L}'$ are just related by a total derivative,
there is a crucial difference between them. While ${\cal L}$ is
gauge invariant, ${\cal L}'$ behaves under gauge transformations
as an operator in the adjoint representation. These transformation 
properties are inherited by $j_A$ and $j'_A$. At the classical level
these currents are conserved and covariantly conserved
respectively \cite{Gracia-Bondia:2000pz,Ardalan:2000qk}
\beq
i) \;\; \partial_\mu j^\mu_A = 0 \;\;\; , \;\;\;\;\;\;\;
ii) \;\; D_\mu j'^{\, \mu}_A=\partial_\mu j'^\mu_A + 
i g \, [A_\mu,j'^{\,\mu}_A]_\ast =0 \, .
\eeq
At the quantum level, $j'_A$ has been shown to satisfy the 
non-commutative counterpart of the ordinary anomaly equation 
\cite{Ardalan:2000cy,Gracia-Bondia:2000pz}. In four dimensions
\beq
D_\mu j'^{\,\mu}_A = - {g^2 \over 16 \pi^2} \epsilon^{\mu \nu \rho \sigma}
F_{\mu \nu} \ast F_{\rho \sigma} \, .
\label{pl}
\eeq
The current $j_A$ cannot satisfy a similar equation,
since $F_{\mu \nu}$ is not gauge invariant. One could think a priori
of two possibilities. The divergence of $j_A$ is equal to some
gauge invariant completion of $F \wedge F$ involving Wilson line
operators. This however would imply that there is an infinite
number of Feynman graphs contributing to the anomaly equation,
in contrast to the ordinary case and to \eqref{pl}. The second 
possibility is that the divergence of $j_A$ remains zero at the
quantum level. Explicit calculations have shown that this is indeed 
the case \cite{Martin:2000qf}, except for a subtlety. 
The previous argument does not constrain those components
of the divergence with zero momentum along the non-commutative directions, 
since $F \wedge F|_{\theta q=0}$ is gauge invariant. 
Indeed, we will show that $j_A$ satisfies
\beq
\int d^2 x_{NC}\,  \partial_\mu j^{\, \mu}_A = 
- {g^2 \over 16 \pi^2} \epsilon^{\mu \nu \rho \sigma}
\int d^2 x_{NC} \, F_{\mu \nu}\, F_{\rho \sigma} \, .
\label{npl}
\eeq
The subindex NC implies integration along the non-commutative plane.
The existence of a restricted anomaly affecting the current $j_A$
was first pointed out in \cite{Ardalan:2000qk}.

Although we have treated the simple case of the axial symmetry, 
the previous considerations apply to any global symmetry and 
to a $U(N)$ gauge group. Let us analyze the implications of 
equation \eqref{npl}. We consider the 
case that $\theta^{0i}=0$, i.e. non-commutativity restricted
to space coordinates. If the cyclic symmetry of the $\ast$-product 
under integration holds, both currents define the same gauge
invariant charge
\beq
Q=\int {\bf dx} \; j^0_A = \int {\bf dx} \; j'^{\, 0}_A \, .
\eeq
The anomaly equation \eqref{pl} implies that $Q$ is not conserved in 
the presence of non-zero instanton number.
If at the same time the divergence of $j_A$ did vanish, we would 
encounter a contradiction. This problem does not arise when 
$j_A$ satisfies instead \eqref{npl}. Then, independently of which
current we use, we will arrive at the same conclusion 
about the variation of the axial charge. 

\section{Perturbative Non-Planar Anomaly Calculation}

\subsection{Two dimensions}

We want to derive the anomaly equation satisfied by the gauge
invariant axial current $j_A$ in two-dimensions. 
We work in Euclidean space, since noncommutativity in the time
coordinate leads to a non-unitary theory \cite{Gomis:2000zz}.
We need to evaluate $\langle j^\mu(q) j_A^\nu(-q)\rangle$, where 
$j^\mu$ denotes the vector current, i.e. $j^\mu=\psi^t {\gamma^\mu}^t 
\ast {\bar \psi}^t$ for a $U(1)$ theory. It is convenient to use the 
relation $\gamma^\mu\gamma^5=i\epsilon^{\mu\nu}\gamma_\nu$, valid in 
two dimensions. This allows us to concentrate on the correlator
$\langle j^\mu(q) j'^{\,\nu}(-q) \rangle$, with $j'^{\,\mu}={\bar \psi} 
\ast \gamma^\mu \psi$. A straightforward and simple calculation gives
\beq
\langle j^\mu(q) 
j'^{\,\nu}(-q)\rangle= - \int {d^2 l \over (2\pi)^2}\ {\rm tr}\ \{
\gamma ^\mu {\slsh{l} \over l^2} \gamma ^\nu {(\slsh{l}+\slsh{q})
\over (l+q)^2} e^{il\cdot\theta\cdot q}\}\label{vp}
\eeq
where the only difference compared to the commutative case is the 
appearance of the momentum-dependent phase factor. Due to the
different order of the fields with respect to the $\ast$-product
in $j$ and $j'$, the correlator of this two operators gives
rise to a non-planar graph (see fig. 1).
\begin{figure}
  \begin{center}
\mbox{\kern-0.5cm
\epsfig{file=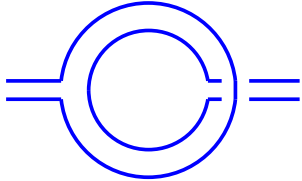,width=4.0true cm,angle=0}}
\label{nonplanar}
  \end{center}
\caption{The non-planar anomaly in 2d.}
\end{figure}

Performing the traces, changing variables and using dimensional 
regularization in \eqref{vp}, one obtains
\bea
&& \langle j^\mu(q) 
j'^{\,\nu}(-q)\rangle = \\
&&= 2\int_0 ^1 dx\int{d^d l\over(2\pi)^d}\ 
{2l^\mu l^\nu-\delta^{\mu\nu}l^2-2x(1-x)q^\mu
q^\nu+\delta^{\mu\nu}x(1-x)q^2\over(l^2+x(1-x)q^2)^2}e^{il\cdot\theta\cdot
q} \nonumber
\eea
For non-vanishing non-commutative momentum $(\theta q)^2$, the 
loop-momentum dependent phase factor renders the integral UV finite and 
dimensional regularization is, in fact, not necessary. 
The integral can be expressed in terms of Bessel functions 
which add up to zero. Thus, for any non-zero $\theta q$, 
the rhs of the previous equation is zero. 
However, at $(\theta q)^2=0$, 
\footnote{In two-dimensional Euclidean space, $(\theta q)^2=0$ implies $q=0$.} 
the non-commutativity parameter does not act as 
a regulator and we have to use dimensional regularization. 
The result for the integral is the same as in the commutative theory
at zero momentum. 

We conclude that $j_A$ is conserved at all non-vanishing 
$\theta q$, {\it i.e.} $q_\mu j_A^\mu=0$, and the non-conservation at 
$\theta q=0$ can be expressed in the form
\beq
\int d^2 x\ \partial_\mu j_A^\mu(x)
=i {g\over 2 \pi}\int d^2 x\ \epsilon^{\mu\nu} F_{\mu\nu}(x) \, .
\label{atwo}
\eeq 
This is an integral equation for the 2d anomaly. Thus we learn that
the anomaly exists and that it is concentrated in the zero momentum 
component of the current. 

\subsection{Four dimensions}

We now repeat the previous argument for $d=4$. Contrary to the
$d=2$ case, we will consider only space non-commutativity. We need to 
evaluate the triangle graph with one insertion of the axial current 
$j_A$. As before, the resulting diagram is non-planar (see 
fig. 2).

\begin{figure}
  \begin{center}
\mbox{\kern-0.5cm
\epsfig{file=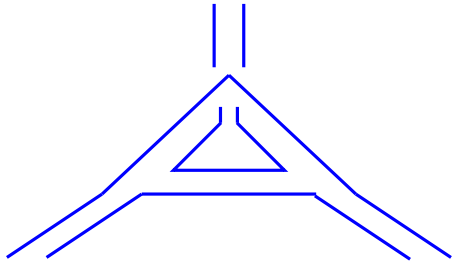,width=6.0true cm,angle=0}}
\label{4d-anomaly}
  \end{center}
\caption{The non-planar anomaly in 4d.}
\end{figure}

In \cite{Martin:2000qf} the anomalies in a $U(N_1) \times U(N_2)$ 
non-commutative gauge theory with a chiral fermion in the
bifundamental representation were treated in great detail. Their 
results imply that if the axial current is coupled through the 
non-planar vertex, then for any non-zero non-commutative momentum
$\theta q$ it satisfies $q_\mu j_A ^{\mu}(q) =0$.
The reason is that for any $\theta q\neq0$, 
due to loop momentum dependent phase factors, the integral no longer 
diverges linearly and the momentum variable can be shifted. 
In particular, the phase factors in the integrals which one gets from 
the two contributing Feynman diagrams coincide after the momentum shift 
and one obtains exact cancellation.

Again, as in the two-dimensional case, this argument breaks down for 
zero external non-commutative momentum, since in
this case the dependence on $\theta$ vanishes and the integrals are not
regulated by non-commutativity. The shift in the integral
now yields a surface term which gives
the anomaly. We then get the known result for the triangle anomaly
for the commutative theory. In summary, the relation \eqref{npl} holds
independent of whether the axial current is coupled via the planar or
non-planar vertex.

In spite of this, it could have seemed tempting to {\it define} 
the value of the non-planar
anomaly diagrams at zero $\theta q$ as the limit from $\theta q$
non-zero. This point of view was followed in 
\cite{Martin:2000qf,Intriligator:2001yu,Martin:2001ye}, which
led to the conclusion that non-planar anomalies cancel. We have argued
that this way of defining correlators leads to inconsistencies,
which are resolved if a non-planar anomaly survives at zero 
non-commutative momentum. This is a surprising conclusion
from the perspective of ordinary quantum field theories, since it
implies that the non-planar anomaly graphs radically violate the 
usual analyticity properties of Green functions. These properties 
are derived from the requirement that commutators of fields vanish for 
space-like separations. However in the non-commutative case this
does not need to hold along the non-local, non-commutative
directions. In particular, it has been shown that
superluminal propagation is possible along the non-commutative 
directions \cite{Landsteiner:2000bw,Hashimoto:2000ys}. 
Therefore, there are no a priori requirements
on the behaviour of quantum correlators as a function of the
variable $\theta q$. On the contrary we should expect an ordinary
behaviour on the variable $q^2_{com}=q_0^2-q_1^2$ 
\cite{Alvarez-Gaume:2001ka}. Notice that we are considering space 
non-commutativity: $[x^2,x^3]=i \theta$. 

It has been known since 
\cite{Minwalla:1999px} that the limit $\theta \rightarrow 0$
of non-commutative Green functions is not smooth. A clear example of 
this is the existence of a non-zero, $\theta$ independent, beta-function 
for pure non-commutative $U(1)$ \cite{Martin:1999aq,Armoni:2000xr}. 
In this paper we are encountering
similar discontinuities of correlators as functions of the
non-commutative momentum. This could have been expected from the fact
that $\theta$ only enters the Feymann integrals in the combination
$\theta q$. 

Without restricting to the particular case studied here, one might wonder 
what is the correct way to define the value of generic correlators at 
zero non-commutative momentum. It would be important to 
carefully study the consequences of a loss of analyticity. 
Another interesting question is the behaviour, with respect to this issue, 
of non-commutative field theories derived from string theory. 
Following this line of ideas, we will discuss in section ~6 the 
non-commutative Green-Schwarz mechanism.

\section{Non-Planar Anomaly via Point Splitting}

The point splitting method provides an alternative and 
neat derivation of the non-planar anomaly. Composite operators 
containing product of fields evaluated at the same point are
generically singular. A proper definition involves a regularization
procedure, which can break some of the symmetries existing at the
classical level. The current $j_A$, regulated by point 
splitting, is 
\beq
j_A^{\mu}(x)={\rm lim}_{\epsilon\rightarrow 0} \;
{\bar \psi} (x + {\scriptstyle {\epsilon \over 2}})\, \gamma^\mu \gamma^5
\ast {\cal U}( x + {\scriptstyle {\epsilon \over 2}}, 
x- {\scriptstyle {\epsilon \over 2}}) \ast
\psi (x - {\scriptstyle {\epsilon \over 2}}) \, 
\label{js},
\eeq
where a Wilson line has been introduced to preserve the gauge 
invariance of the regularized expression,
\beq
{\cal U}(x, y) = e^{i g \int_x^y dl\cdot A(l)}  \, .
\eeq
The divergence of $j_A^{\mu}$ can be derived by expanding the 
Wilson line in \eqref{js} at linear order in $\epsilon$ and using 
the equations of motion of the fermionic fields \cite{Ardalan:2000cy}
\beq
\partial_\mu j_A^{\mu}(x)=-i g \, {\rm lim}_{\epsilon \rightarrow 0} \;
\epsilon^\nu
{\bar \psi} (x + {\scriptstyle {\epsilon \over 2}})\, \gamma^\mu \gamma^5
\ast F_{\mu \nu} (x) \ast
\psi (x - {\scriptstyle {\epsilon \over 2}}) \, \label{djs}.
\eeq

The rhs of this equation, although proportional to $\epsilon$, may give a 
finite result since the correlator of the fermion fields can become 
singular as the regulator, $\epsilon$, is removed.
Namely, in Fourier space we will have
\bea
& i q_\mu j_A^{\mu}(q) & = -i g {\rm lim}_{\epsilon \rightarrow 0}  
 \epsilon^\nu \int {d^d l \over (2 \pi)^d}{d^d p\over (2 \pi)^d} \;  
e^{{i \over 2}(l-p)\epsilon } e^{-{i \over 2}(l \theta q+ q \theta p 
+ p \theta l)}
\;\;\;\;\;\;\; \nonumber \\
&& \;\;\;\;\;\;\;\;\;\;\;\;\;\;\;\;\;\;\;\;\;\;\;\;\;\;\;\;\;\;\;\;\; 
\langle \, {\bar \psi} (l)\, \gamma^\mu \gamma^5 \psi(p) \, \rangle \,
F_{\mu \nu} (q-l-p) \, ,
 \label{npaF}
\eea
with $d$ the space-time dimension. We will treat first the 
two-dimensional case \footnote{Here we also assume an Euclidean space.}.
Substituting the tree level fermion correlator and performing the $p$ 
integration we obtain
\bea
i q_\mu j_A^{\mu}(q)& =& i g \; {\rm lim}_{\epsilon \rightarrow 0}  
 \epsilon^\nu F_{\mu \nu} (q) \, {\rm tr} (\gamma^\rho \gamma^\mu \gamma^5) 
\int {d^2 l \over (2 \pi)^2} 
{l_\rho \over l^2} e^{i l (\epsilon - {\theta q})} 
\nonumber \\
&=&i {g \over \pi} \;  {\rm lim}_{\epsilon \rightarrow 0} 
{\epsilon^\nu (\epsilon - \theta q) _\rho \over (\epsilon-{\theta q})^2}\; 
\epsilon^{\mu \rho} F_{\mu \nu}(q) \, .
\label{npa}
\eea
When ${\theta q} \neq 0$, the rhs of this expression tends to zero. 
We recover thus the result that 
non-planar anomalies cancel. The order in which the fermion fields and 
the field strength are multiplied in \eqref{djs} is at the origin of
this cancellation. It causes that, even after sending $\epsilon$ to 
zero, the fermion fields are effectively separated by a distance 
$\theta q$, being $q$ the momentum carried by the field strength.
Therefore their correlator does not become singular.  
This makes clear why the cancellation is not operative
at ${\theta q}=0$. Indeed, at this value \eqref{npa} gives a 
non-zero result, as it is the 
case for ordinary theories. In dimension $d$ we have
${\rm lim}_{\epsilon \rightarrow 0} 
{\epsilon^\nu \epsilon_\rho \over \epsilon^2}= 
{\delta_\rho^\nu \over d}$. Substituting this in \eqref{npa} we obtain 
that the current $j_A^{\mu}$ satisfies the integrated form of the 
standard anomaly, equation \eqref{atwo}. This fact is implicit in
the analysis of \cite{Ardalan:2000cy}. 

The anomaly of the non-gauge invariant axial current $j'_A$ in
\eqref{currents} can be studied in the same way. The regularized version 
of the current is in this case
\beq
{j'}_A^{\mu}(x) = -\, {\cal U}( x , x + {\scriptstyle {\epsilon \over 2}})
\ast \psi^t (x + {\scriptstyle {\epsilon \over 2}})\, (\gamma^\mu \gamma^5)^t
\ast {\bar \psi}^t (x - {\scriptstyle {\epsilon \over 2}})
\ast {\cal U}( x - {\scriptstyle {\epsilon \over 2}}, x)  \, .
\label{pjs}
\eeq
The Wilson lines have been introduced in order that the point splitting
regularization does not alter the behavior of the current under
gauge transformations. The divergence of ${j'}_A^{\mu}$ is 
\bea
\partial_\mu {j'}_A^{\mu}(x)& =& -{i g \over 2} \, 
{\rm lim}_{\epsilon \rightarrow 0} \epsilon^\nu \left[
F_{\mu \nu} (x) \ast \psi^t (x + 
{\scriptstyle {\epsilon \over 2}})\, (\gamma^\mu \gamma^5)^t
\ast {\bar \psi}^t (x - {\scriptstyle {\epsilon \over 2}})
+ \right. \nonumber \\
&& \;\;\;\;\;\;\;\;\;\;\;\;
+ \left. \psi^t (x + {\scriptstyle {\epsilon \over 2}}) \, 
(\gamma^\mu \gamma^5)^t \ast {\bar \psi}^t 
(x - {\scriptstyle {\epsilon \over 2}})
\ast F_{\mu \nu} (x) \right] 
\eea
It is by now well known that the current $j'_A$ satisfies
the non-commutative counterpart of the usual anomaly equation of 
gauge theories \cite{Ardalan:2000cy,Gracia-Bondia:2000pz}. We can 
easily see how this result
arises in the point splitting approach. The crucial difference
between the equations for the divergence of $j_A$ and $j'_A$ is the order
in which the field strength and the fermionic fields are multiplied.
In \eqref{pjs} the field strength does not separate $\psi$ and 
${\bar \psi}$ and the effective regularization of the current due to 
the star-product does not take place. Namely, the divergence
of $j'_A$ is given by equation \eqref{npa} with the argument of the 
$l$-integration being just $\epsilon$ instead of $\epsilon - {\theta q}$.

We will evaluate now the divergence of the non-planar current $j_A$
in four dimensions using the same approach. The main difference with 
the two-dimensional case is that the relevant contributions 
to the fermion correlator in \eqref{npaF} 
come from first and second order in perturbation theory. The 
contribution from tree level cancels 
since ${\rm tr} \big( \gamma^\rho \gamma^\mu \gamma^5 \big)\! =\!0$ in 
$d\!>\!2$. At first order in perturbation theory we have
\beq
-ig\langle \, {\bar \psi}(l) \gamma^\mu \gamma^5 \psi (p) 
\int d^4 x \; {\bar \psi} \, \ast \! \slsh{\! A} \ast \psi \, \rangle =
\,-4 g \, \epsilon^{\mu \rho \alpha \beta} \,  
{l_\rho p_\alpha \over l^2 p^2} \, A_\beta(p+l) \; 
e^{{i\over 2} p\theta l} \, ,
\eeq
where we have used that in four dimension 
${\rm tr} \big( \gamma^\mu \gamma^\rho \gamma^\alpha 
\gamma^\beta \gamma^5 \big)\! =\!-4\,i\,
\epsilon^{\mu \rho \alpha \beta}$. 
Defining $k\!=\!p\!+\!l$ and substituting the previous
equation in \eqref{npaF}, we obtain
\bea
&& \!\!\!\!\!\!\!\!\! i q_\mu j_A^{\mu}(q)  =  \label{fourdj} \\
&& \!\!\!\!\!\!=4i g^2\; {\rm lim}_{\epsilon \rightarrow 0}  
\epsilon^\nu \epsilon^{\mu \rho \alpha \beta}  
\int {d^4 k \over (2 \pi)^4}  {d^4 l \over (2 \pi)^4}
\, {l_\rho \,e^{i l (\epsilon -
{\theta q})} \over l^2 (k-l)^2} \,
F_{\mu \nu} (q-k) \; k_\alpha A_\beta(k) \, e^{-{i \over 2} q \theta k}  
\nonumber \\
&& \!\!\!\!\!\! = - {g^2 \over 2 \pi^2} \; {\rm lim}_{\epsilon \rightarrow 0}  
{\epsilon^\nu (\epsilon \!-\!
{\theta q})_\rho \over (\epsilon\!-\!{\theta q})^2}\; 
\epsilon^{\mu \rho \alpha \beta}  \int {d^4 k \over (2 \pi)^4}\;
f(k\!,\!|\epsilon\! - \!{\theta q}|)\;F_{\mu \nu} (q-k) \;i k_\alpha
A_\beta(k) \, e^{-{i \over 2}q \theta k} \nonumber \, .
\eea
The $l$-integration in this case produces, in addition to 
$(\epsilon\!-\!{\theta q})_\rho/(\epsilon\!-\!{\theta q})^2$, 
the function $f(k,|\epsilon\! - \!{\theta q}|)$. $f$ 
can be expressed in terms of Bessel functions, it is always finite
and tends to $1$ as $|\epsilon\! - \!{\theta q}|$ 
tends to zero. Thus the rhs of \eqref{fourdj} vanishes for
${\theta q} \neq 0$, as it happens in the two-dimensional case. 
For ${\theta q}=0$ it gives clearly a non-zero result. It is 
straightforward to analyze the contribution to the anomaly equation 
from evaluating the fermion correlator at second order in 
perturbation theory. We will not do it explicitly, since the pattern is
as before. At ${\theta q}\neq 0$ the anomaly cancels. The result 
at ${\theta q}=0$ is non-zero, combining with  \eqref{fourdj}
to promote $i (k_\alpha A_\beta(k) - k_\beta A_\alpha (k))
\rightarrow F_{\alpha \beta}(k)$. Restricted to ${\theta q}=0$,
\eqref{fourdj} reduces to
\beq
i q_\mu j_A^{\mu}(q) =-{g^2 \over 16 \pi^2} \;
\int {d^4 k \over (2 \pi)^4}\; \epsilon ^{\mu \nu \alpha \beta}F_{\mu \nu} 
(q-k) \; F_{\alpha \beta}(k)  \, ,
\eeq
which reproduces the integrated anomaly 
equation \eqref{npl}.

\section{Mixed Anomalies}

Similar arguments apply to the local anomalies of chiral gauge 
theories. Consider a simple case among those treated in \cite{Martin:2000qf},
a $U(1) \times U(1)$ gauge theory with a chiral fermion in the bifundamental 
representation. In the case of a right handed fermion, the vector currents 
associated to the groups $U(1)_1$ and $U(1)_2$ are
\beq
 j_1^{\mu} = \psi^t ({\gamma^\mu} P_+)^t \ast {\bar \psi}^t
\;\;\; , \;\;\;\;\;\;
 j_2^{\mu} = {\bar \psi} \ast (\gamma^\mu P_+) \psi \, ,
\label{currents2}
\eeq
with $P_+={1 \over 2} (1\! +\! \gamma^5)$. These two currents are in one 
to one correspondence with those in \eqref{currents}. Thus the
analysis of previous sections applies straightforwardly
to local mixed anomalies.
We will reduce now to the four dimensional case and only spatial
non-commutativity. Equation \eqref{npl} implies 
the existence of a mixed $U(1)_i U(1)^2_j$ anomaly restricted to
$\theta q=0$. Suppose that we add the necessary chiral matter to cancel
$U(1)^3$ anomalies, leaving the mixes anomalies untouched. Then we
have
\beq
\int d x_{NC}\,  \partial_\mu j^{\, \mu}_{(i)} = 
- {1 \over 32 \pi^2} \epsilon^{\mu \nu \rho \sigma}
\int d x_{NC} \, F_{\mu \nu}^{(j)}\, F_{\rho \sigma}^{(j)} \, ,
\label{nplm}
\eeq
for $i \neq j$. The variation of the action induced by a gauge 
transformation $\lambda^{(i)}$ due to the anomaly \eqref{nplm} is
given in Fourier space by
\beq
\delta_{\lambda^{(i)}} S = {1 \over 16 \pi^2 } \int d^4 q\ 
\lambda ^{(i)}(q) F \tilde F^{(j)} (-q) | _{\theta q=0}\,. \label{prepain}
\eeq
The variation \eqref{prepain} vanishes unless the gauge parameter
is a delta function $\lambda (q) \propto \delta (\theta q)$. In
coordinate space the action \eqref{prepain} takes the form
\beq
\delta_{\lambda^{(i)}} S = {1 \over 16 \pi^2 V_{NC}} \int d^2 x 
\left( \int d^2 x_{NC}\, \lambda^{(i)} \right) \left( \int d^2 x_{NC}\, 
F {\tilde F}^{(j)} \right)  \, , 
\label{pain}
\eeq
where $V_{NC}$ represents the (infinite) volume of the non-commutative
directions, and $x$ and $x_{NC}$ denote respectively the ordinary and
non-commutative directions. The variation of the action is non-zero only 
if the integration of $\lambda$ along the non-commutative directions
would cancel the volume factor in the denominator of \eqref{pain}.

\section{The Noncommutative Green-Schwarz Mechanism}

In this section we address the paradox of the
${\mathbb C}^3/{\mathbb Z}_3$ orbifold, 
which was mentioned in the introduction 
and which was raised in \cite{Intriligator:2001yu}.
Consider a collection of $N$ D3 branes, placed on a
${\mathbb C}^3/{\mathbb Z}_3$ orbifold
singularity in a background of a constant NS-NS 2 form. The resulting 
noncommutative field theory is a $U(N)^3$ field theory
with three generations of bi-fundamental chiral multiplets. The 
matter spectrum is summarized in table 1.

 \begin{table}
\begin{displaymath}
\begin{array}{l c@{ } c@{ } c@{ } c}
 & \multicolumn{1}{c@{\times}}{U_1(N)}
& \multicolumn{1}{c@{\times}}{U_2(N)}
& \multicolumn{1}{c@ {}}{U_3(N)} \\
\hline
3\ \rm {chirals} & \Yfund & \overline{\Yfund}  & 1 \\
3\ \rm {chirals} & 1 & \Yfund & \overline{\Yfund}  \\
3\ \rm {chirals} & \overline{\Yfund}  & 1 & \Yfund \\
\end{array}
\end{displaymath}
\caption{The matter content of the ${\mathbb C}^3/{\mathbb Z}_3$ 
orbifold theory.}
\label{content}
\end{table}

Since the theory is chiral there are potential anomalies. The
$U_i(N)^3$ anomalies cancel, since there are $3N$ fundamental and
$3N$ anti-fundamental fermions charged under each of the gauge groups factors.
There is, however, a potential $U_i(N)^2 U_j(1)$ anomaly.
For example, the first row in table 1 describes the matter content which is
charged under both $U_1(N)$ and $U_2(N)$. Only these fermions
circulate in a triangle diagram with two external legs in 
$U_1(N)$ and one leg in $U_2(N)$. One can choose a regularization
scheme in which the local $U_2(1)$ gauge symmetry is anomalous. Hence,
the field theory is sick 
(to be more precise, only two combinations out of the
three $U(1)$'s are anomalous).
Since this theory arises from a string theory setup, it should be
consistent - probably even in the decoupling limit $\alpha' \rightarrow 0$

Let us review first the string theory solution to this problem in the
commutative case \cite{Ibanez:1998qp}. The type IIB
closed string spectrum contains two massless   
0-forms fields in the twisted sectors which are localized at the
orbifold singularity. Their coupling to the
brane can be written in a convenient way in terms of three fields $C^{(i)}$.
They transform under gauge transformations 
$\delta({\rm tr}A_\mu^{(i)})=\partial_\mu\epsilon^{(i)}$ and
$\delta C^{(i)}=-\epsilon^{(i)}$. The $C^{(i)}$ 
are constraint to sum to zero in a gauge invariant way. The action is  
\bea
& & 
S \sim \int d^4 x\ \left \{ {1\over \alpha'} \sum _{i=1} ^3 \left ( ({\rm tr}\
  A^{(i)}_\mu + \partial _\mu C^{(i)})^2 + \lambda (\partial ^2
  C^{(i)}
 + {\rm tr}\ \partial A^{(i)}) \right ) \right . \nonumber \\
& &
\left . \, \, \, \, \,\,\,\,\,\,\,\,
+\left (
 C^{(1)} \, ({\rm tr}\ F_{\mu \nu}^{(2)} \tilde F_{\mu \nu}^{(2)}
-{\rm tr}\ F_{\mu \nu}^{(3)} \tilde F_{\mu \nu}^{(3)}) + {\rm cyclic\ perm.}
\right ) \right \}\,.
 \label{GSaction}
\eea
Upon integration over the three twisted RR-fields and the Lagrange multiplier
$\lambda$ we get the effective action 
\beq
S_{eff} =  {3\over 16\pi^2}\int d^4 x \left\{ ({\rm tr}\ \partial\! \cdot \! 
A ^{(1)})
{1\over \partial ^2} \left({\rm tr}\ F\tilde F^{(2)} - {\rm tr}\ F\tilde
F^{(3)}\right) + {\rm cyclic\ perm.} \right\} . \label{Seff}
\eeq
In addition there is a rank two mass matrix for the three $A_\mu^{(i)}$'s 
and also $(F\tilde F)^2$ terms. The full partition function is gauge invariant. Under a $U_i(1)$ gauge
rotation, the anomaly is compensated by the gauge transformation of 
\eqref{Seff}.
In a diagrammatic language it is described by fig. 3. 
\begin{figure}
  \begin{center}
\mbox{\kern-0.5cm
\epsfig{file=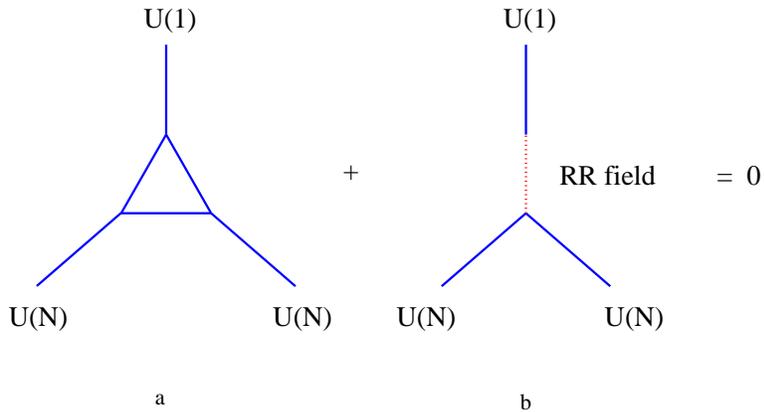,width=10.0true cm,angle=0}}
\label{GS}
  \end{center}
\caption{The Green-Schwarz mechanism.}
\end{figure}
The triangle diagram (fig. 3a) is cancelled by a tree level diagram which
involves an exchange of closed RR field (fig. 3b), in a generalized 
Green-Schwarz mechanism. The diagram, fig. 3b, can be understood as due to 
an annulus diagram, with two $U(N)$ insertions on one boundary and one
$U(1)$ insertion at the second boundary (see fig. 4).  
\begin{figure}
  \begin{center}
\mbox{\kern-0.5cm
\epsfig{file=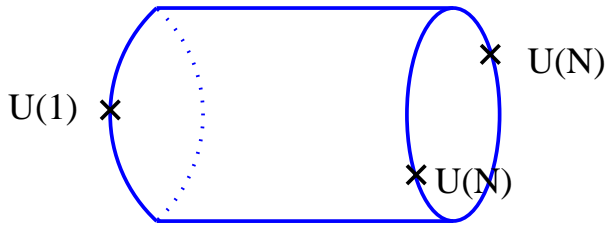,width=8.0true cm,angle=0}}
\label{annulus}
  \end{center}
\caption{The annulus diagram with $U(N)^2$ insertions on one boundary
  and $U(1)$ insertion at the other.}
\end{figure}
Thus, in the commutative case gauge invariance
is restored by an axion like field (twisted RR field) that cancels
the gauge anomaly and gives a gauge invariant mass $M^2 \sim {1\over
\alpha'}$ to the two anomalous $U(1)$'s. The third $U(1)$ remains massless.
The low energy theory, in the commutative case, would therefore be an
$SU(N)^3\times U(1)$ gauge theory. The two apparently anomalous local
$U(1)$'s became global and decoupled, yielding a consistent
low-energy theory.

Let us now return to the more complicated non-commutative case. 
We will consider again the case where the non-commutativity affects 
only (two) space
directions. Notice that the ${\mathbb C}^3/{\mathbb Z}_3$ 
orbifold gauge theory has ${\cal N}=1$ supersymmetry. Thus it
is free from the dangerous quadratic infrared divergences due 
to UV/IR mixing effects, which could render the theory unstable
\cite{Matusis:2000jf,Ruiz:2000hu,Landsteiner:2001ky}. However,
 we would like to stress that the triangle 
anomaly does not vanish, due to the contribution from 
zero non-commutative momentum flowing into the triangle
diagram at the non-planar vertex. This is exactly the non-planar 
anomaly which was described in the previous sections. Moreover, the
problem is more severe in the noncommutative case, since $U(1)$
couples to the rest of the $SU(N)$ theory due to non-commutative 
gauge invariance \cite{Armoni:2000xr} and therefore the problem cannot
be solved by making the $U(1)$ global (massive). 

Furthermore, the action \eqref{GSaction} now presents some immediate
problems. We will assume that if a Green-Schwarz mechanism exists in 
the non-commutative case, it must be possible to formulate it 
naturally in terms of the non-commutative gauge field $A_\mu$.
The reason for this is that the non-planar anomaly has a simple 
expression in terms of $A_\mu$, instead of involving arbitrary powers 
of this field. However, the first term in \eqref{GSaction} would 
not be gauge invariant under non-commutative gauge transformations.
Second the non-planar anomaly, diagram 3a, exits only for zero non-commutative 
momentum $\theta q=0$. However the diagram 3b mediated by the action 
\eqref{GSaction} seems to exist also for non-zero $\theta q$. 

In spite of these problems, it is easy to propose an action
similar to \eqref{Seff} whose variation under gauge transformations
would cancel the non-planar anomaly
\beq
S_{eff} = {3 \over 16 \pi^2 V_{NC}}
\int d^2 x \left\{ ({\rm tr}\ \partial \! \cdot \! A^{(1)})_0
\, {1\over \partial ^2} \, ({\rm tr}\ F \tilde F^{(2)} - {\rm tr}\ F \tilde
F^{(3)})_0 + {\rm cyclic\ perm.} \right\} \, , \label{GSNC}
\eeq
where, as in section 5, $V_{NC}$ is the volume of 
and the subscript ``$_0$'' denotes integration along the non-commutative 
plane. The coordinates $x$ as well as the derivatives in \eqref{GSNC}
are restricted to the two commutative directions. It is immediate that
the gauge variation of the previous expression can cancel the variation
of the action due to the non-planar anomaly, eq. \eqref{pain}.  
The action \eqref{GSNC} is well defined under {\em noncommutative}
gauge transformation since each trace operator is separately integrated
along the non-commutative plane.
If we interpret \eqref{GSNC} as derived from a diagram
as in fig 3b, it would correspond to restricting the closed
string exchange to $\theta q=0$. Notice that only RR fields from the
twisted sectors can participate in the Green-Schwarz mechanism.
In the ${\mathbb C}^3/{\mathbb Z}_3$ orbifold there are only
two twisted massless 0-form RR fields, which should couple to
a combination of $U(1)$'s as in the ordinary case. Although 
a sum of two $U(1)$'s is {\em not} a noncommutative $U(1)$, 
one is allowed to consider linear combinations at $\theta q=0$,
where the non-commutative structure trivializes. 

We now indicate how the action \eqref{GSNC} results from
string theory. Consider the annulus diagram, fig. 4. The
calculation of this diagram for the noncommutative theory, based on
the expression for the ordinary theory \cite{Green:mn}, was carried out in
\cite{Intriligator:2001yu}
\beq
A^{\mu\nu}\propto\epsilon^{\mu\nu\rho\sigma}p_\rho k_\sigma\lim _{M\rightarrow \infty} 
M^2 \int _0 ^\infty dt\ e^{-M^2 t}
e^{-{\pi \over 2t} (\theta q)^2} \int _0 ^1 d\nu_{1,2} 
e^{-{i\over 2} p\theta k (2\nu _{12} + \epsilon (\nu _{12}))}, \label{KI}
\eeq
where $M$ is a regulator, $\epsilon$ is the step function and $\nu
_{12} = \nu_1 - \nu_2$.
For any non-zero $\theta q$ the integral near $t=0$ (the UV) is exponentially
suppressed and yields a zero anomaly. However, for $\theta q=0$ this is
no longer true and we have an anomaly. The interpretation is that for 
$\theta q =0$ we have an anomaly in field theory (non-vanishing
triangle diagram) and it is compensated 
in a Green-Schwarz mechanism by an exchange of a massless RR field.
The mechanism for $\theta q=0$ is exactly the same as in the commutative
theory.

For $\theta q\ne 0$ the scenario is different. The amplitude \eqref{KI} 
vanishes in this case, indicating that there is no anomaly in the 
field theory. 
The crucial difference with respect to the previous case is that, in
considering a closed string exchange as in fig. 3b, we cannot
discard the massive string modes \cite{Arcioni:2000bz}. The on-shell
 condition for twisted
closed string modes is $g^{\mu \nu}q_\mu q_\nu + {N\over \alpha'} =0$, with $g$
denoting the closed string metric. The open and closed string metrics 
are related via  \cite{Seiberg:1999vs}
\beq 
g^{-1} = G^{-1} - {\theta G \theta \over (2\pi \alpha')^2} \label{Cmetric},
\eeq
and thus the on-shell condition is
\beq
q^2 + {(\theta q)^2 \over (2 \pi \alpha')^2} = - {N \over \alpha'}.
\label{disp}
\eeq
In the $\alpha' \rightarrow 0$ limit, the oscillator mass becomes a subleading
effect with respect to the momentum along the non-commutative directions.
Assuming that the massive RR string modes have similar couplings to the
field theory operators as the massless one, it is clear that all
of them will contribute to diagram 3b when $\theta q\ne 0$.
 Such an exchange is 
then more adequately interpreted directly in field theory terms. Its effect is
to introduce the non-commutative damping factor $e^{-{\pi \over 2t} (\theta q)^2}$
that leads to the cancellation of \eqref{KI}. Notice that
in the $\alpha' \rightarrow 0$ limit, modes with $\theta q \ne 0$ are
never on-shell because the kinetic term is suppressed by two powers of $\alpha'$
with respect to the momentum in the non-commutative directions.
Contrary, when $\theta q= 0$
the oscillator mass is the dominant effect in \eqref{disp}. This selects
the massless closed string mode, bringing us back to the ordinary
Green-Schwarz mechanism.

The Green-Schwarz mechanism has its origin in the fact that non-planar
string annulus diagrams are generically finite. The absence of divergences in 
the non-planar annulus can be understood in terms of the closed string
modes using channel duality. Consequently, closed string modes play a 
fundamental role in the Green-Schwarz mechanism \cite{Green:sg}. 
The (partial) cancellation of the non-planar anomaly in non-commutative
gauge theories occurs because of the regularization of the associated
triangle graph by non-commutative effects. This regularization 
of otherwise divergent Feymann graphs gives
rise to one of the most remarkable characteristics of non-commutative
field theories: UV/IR mixing \cite{Minwalla:1999px}. In
\cite{Minwalla:1999px,Liu:2000ad,Kiem:2001fn} the
similarity between the non-planar graphs in non-commutative field
theories and in string theory was stressed. Following
these ideas, \cite{Armoni:2001uw} showned that 
the infrared divergences associated with UV/IR mixing effects in gauge 
theories can be reproduced in terms of closed string exchange, where
the whole tower of closed string modes contributes. 
We propose that a similar pattern applies to the cancellation of non-planar 
anomalies (see also \cite{Intriligator:2002ii} for a related discussion 
about flavour anomalies). In this sense, it can be interpreted as 
an automatic Green-Schwarz mechanism \cite{Intriligator:2001yu}.

Since at $\theta q = 0$ we have argued that an ordinary Green-Schwarz 
mechanism takes place, we propose the following Lagrangian for the
`mass term' of the $U(1)$ in the noncommutative theory
\beq
{\cal L} \sim {1\over \alpha'} ({\rm tr}\ A_\mu + \partial _\mu
C)(q) G^{\mu \nu}({\rm tr}\ A_\nu + \partial _\nu
C)(-q) |_{\theta q=0}. \label{GSmass}
\eeq
The Lagrangian \eqref{GSmass} is gauge invariant with respect to
noncommutative gauge transformations. Now, we add to the action the 
'axion' term $C F \tilde F$ and we integrate over the RR field. The 
resulting action is \eqref{GSNC}.

There is a subtle issue that should be discussed. The WZ
couplings in the noncommutative theory are more involved than in the
ordinary case, see details at \cite{Mukhi:2000zm,Liu:2001pk}.
However, we assume that the existence of a similar piece at zero
noncommutative momentum is enough for the anomaly cancellation.   

We have argued that the non-commutative gauge theory on the 
${\mathbb C}^3/{\mathbb Z}_3$ orbifold has non-vanishing 
mixed anomalies. However they only affect $U(1)$ modes
with zero non-commutative momentum. For these modes
both the non-commutative and the non-abelian structure 
trivialize, i.e. they do not mix with the rest under
gauge transformations. In \eqref{GSmass} we have proposed
that due to the interaction with the RR field, the
$\theta q=0$ components of two of the three $U(1)$'s
become massive. A linear combination of $U(1)$'s remains
anomaly free as in the ordinary case. Thus the non-commutative
$U(N)^3$ structure is preserved, since this only requires 
that the $U(1)$ excitations with non-zero noncommutative
momentum remain coupled. The excitations with zero
noncommutative momentum become massive and decouple.
It is interesting that this
splitting, which of course violates Lorentz symmetry, preserves the
full gauge symmetry.

\section{Global Anomalies}
In this section we address the issue of global
anomalies. As we have already mentioned in the introduction, there are
two different kinds of anomalies in noncommutative theories: planar
and nonplanar. In particular there are two kinds of global anomalies.
Since planar global anomalies are as in the
commutative situation, we will only consider nonplanar anomalies. 

Consider a fermion that transforms in the bifundamental 
representation of a product
group $G\times H$, where $H$ is a global symmetry  
and $G$ is either local or global. Both cases have interesting applications in 
particle physics. 

The prime example of mixed global and local symmetries is the $\pi ^0
\rightarrow 2\gamma$ process. In this case the global symmetry is $U(1)_A$ and
the local is $U(1)$ (Maxwell). According to the analysis in the
previous sections the axial charge is not conserved. Thus, though
we cannot write a local anomaly equation, we see that no conservation
of charge forbids the pion decay, in contrast to \cite{Intriligator:2001yu}. 

The second important case is when both $G$ and $H$ are global. Here
 the global anomaly restricts the microscopic structure of the theory
(or the description in terms of a dual theory, as in
 \cite{Seiberg:1994pq}). According to 't Hooft \cite{tHooft}, the anomaly,
 which is nothing but the number of fermions
 species circulating in the triangle loop, should be the same in the
 two descriptions (see also\cite{Frishman:1980dq}).
 This condition imposes stringent restrictions on a
 possible dual description. 

 Had the nonplanar anomaly vanished, this restriction could have been
 removed. However, we claim that the same anomaly matching conditions
 as in the commutative theory should be imposed. To see that, we can
 repeat the reasoning of 't Hooft. One can gauge the global symmetry
 with a very weak coupling 
\footnote{We assume here a global symmetry that has a natural
realization in terms of a $U(N)$ group.
Also, in order to have any potential local anomaly we must work 
in more than two dimensions with some commutative directions.}
and add chiral matter to cancel the anomaly.
 The same 'hidden' chiral sector should guarantee an
 anomaly free gauge theory in the two descriptions. Therefore the
 global anomaly should be the same. For this kind of
 reasoning in favor of the anomaly matching
 conditions it is enough that an integral version of the
 anomaly exist, since even an integral version of the anomaly renders
the local theory inconsistent. Thus, the noncommutative theory is not less
 restrictive than the commutative one.

\Acknowledgements

We would like to thank L. Alvarez-Gaume, C. Angelantonj, F.
Ardalan, O. Bergman, S. Elitzur, A. Font, Y. Frishman, K. Landsteiner,
C. Martin, R. Minasian, R. Rabadan, C. Scrucca, S. Silva, 
M. Vazquez-Mozo, Y. Oz, A. Uranga and F. Zamora for discussions.
We would like to thank A. Schwimmer for his comments on the 
first version of the paper. They led to the additional discussion 
at the end of sect.~3.2.
S.T. would like to thank the Isaac-Newton-Institute for Mathematical
Sciences for hospitality during the final stages of this work.
A.A. thanks the Albert-Einstein-Institut for the warm hospitality
where part of this work was done. This work was
 partially supported by GIF, the German-Israeli Foundation 
for Scientific Research.

\end{document}